\documentclass[1p,sort&compress]{elsarticle}
\usepackage{amsmath, amssymb}
\usepackage{color}

\newcommand{\qmA}{\hat{A}}
\newcommand{\qmB}{\hat{B}}

\newcommand{\qmH}{\hat{H}}

\newcommand{\qmQa}{\hat{Q}^{[a]}}
\newcommand{\clQa}{Q^{[a]}}
\newcommand{\qmQb}{\hat{Q}^{[b]}}

\newcommand{\qmrho}{\hat{\rho}}

\newcommand{\Tr}{\mathrm{Tr}}
\newcommand{\hdelta}[1]{\hat{\delta}_{#1}}
\newcommand{\hL}[1]{\hat{L}_{#1}}
\def\<#1>{\langle #1 \rangle}

\begin{document}
\begin{frontmatter}

  \title{Derivation of the density operator with quantum analysis for the generalized Gibbs ensemble in quantum statistics}
  \journal{}
  \author{Masamichi Ishihara\corref{cor}}
  \cortext[cor]{Corresponding author. Tel.: +81 24 932 4848; Fax: +81 24 933 6748.}
  \ead{m\_isihar@koriyama-kgc.ac.jp}
  \address{Department of Human Life Studies, Koriyama Women's University, Koriyama, Fukushima, 963-8503, JAPAN}
  \begin{abstract}
We derived the equation of the density operator for generalized entropy and generalized expectation value 
with quantum analysis when conserved quantities exist. 
The derived equation is simplified when the conventional expectation value is employed.
The derived equation is also simplified
when the commutation relations, $[\qmrho, \qmH]$ and $[\qmrho, \qmQa]$, are the functions of the density operator $\qmrho$, 
where $\qmH$ is the Hamiltonian, and $\qmQa$ is the conserved quantity.
We derived the density operators for the von Neumann entropy, the Tsallis entropy, and the R\'enyi entropy
in the case of the conventional expectation value.
We also derived the density operators for the Tsallis entropy and the R\'enyi entropy
in the case of the escort average (the normalized $q$-expectation value), 
when the density operator commutes with the Hamiltonian and the conserved quantities. 
We found that 
the argument of the density operator for the canonical ensemble is simply extended to the argument for the generalized Gibbs ensemble
in the case of the conventional expectation value, even when conserved quantities do not commute.
The simple extension of the argument is also shown in the case of the escort average, 
when the density operator $\qmrho$ commutes with the Hamiltonian $\qmH$ and the conserved quantity $\qmQa$:
$[\qmrho, \qmH] = [\qmrho, \qmQa]=0$.
These findings imply that 
the argument of the density operator for the canonical ensemble is simply extended to 
the argument for the generalized Gibbs ensemble in some systems.
\end{abstract}


\end{frontmatter}

\section{Introduction}
The density operator is widely used to calculate physical quantities in various systems. 
Therefore, it is required to determine the density operator with adequate constraints. 
A method to obtain the density operator with constraints is the maximum entropy principle (MEP). 
The maximum entropy principle is useful to determine the density operator 
under constraints when an adequate entropy is given. 
The density operators for some entropies were derived in the MEP \cite{Jaynes-PR106}. 
A density operator is applied to calculate values of physical quantities. 

The generalized Gibbs ensemble (GGE) \cite{Wang1983, Rigol2007, Calabrese2015} is an extension of the canonical ensemble, 
and some conserved quantities are contained in the density operator for the GGE.
The density operator for the GGE was derived in the MEP \cite{Rigol2007, Langen-Science2015}, 
and it is shown that the argument of the density operator is simply extended for the von Neumann entropy with the conventional expectation value. 
It is considered that some systems can be described by the GGE.

A nonextensive statistics is an extension of the Boltzmann-Gibbs statistics, 
and many entropies have been proposed. 
For example, the Tsallis \cite{Tsallis1998, Book:Tsallis, Bercher2011, Pougaza:AIPConf:2011}
and the R\'enyi  entropies \cite{Bercher2011, Pougaza:AIPConf:2011, Abe-PRE94}
are well-known entropies.
A difference between the Boltzmann-Gibbs statistics and the unconventional statistics is the definition of the expectation value.  
The escort average (the normalized $q$-expectation value) is often used in the statistics.
The density operator can be obtained in the MEP, and is applied to various phenomena.  
The constraint of energy conservation is imposed in such calculations. 
It is natural to consider some additional constraints of conserved quantities. 
 

The quantum analysis developed by M.~Suzuki 
\cite{Suzuki-commun-math, Suzuki:Book:QuantAnalysis, Suzuki-Review-MathPhys, Suzuki:JMathPhys, Suzuki-progress, Suzuki-IJMPC10}
is a useful tool to obtain the density operator in the MEP. 
The quantum analysis was applied to
the nonequilibrium response \cite{Suzuki-progress} and quantum correlation identities \cite{Suzuki-IJMPC10}.
The density operators for some entropies were derived with the quantum analysis in the previous study \cite{Ishihara-physicaA}. 
It is expected that the quantum analysis works well in order to obtain the density operators with some constraints in the MEP.


In this paper, we attempt to obtain the density operators with the quantum analysis in the MEP. 
Some entropies such as Tsallis and R\'enyi entropies are extremized under the existence of conserved quantities.
Density operators are explicitly obtained in some cases, 
and conditions to obtain the well-known density operators for the GGE are explicitly shown. 
The density operators in some cases are given with the quantum analysis in the MEP.

We derived the equation of the density operator for generalized entropy and generalized expectation value
when conserved quantities exist.
It was shown that the equation of the density operator is simplified in particular cases, 
and the density operators for the von Neumann, Tsallis and R\'enyi entropies were obtained.  
We found the followings.
The argument of the density operator for the canonical ensemble is simply extended to the argument for the GGE
without assuming commutation relations between conserved quantities, when the conventional expectation value is employed. 
The extension is possible for the unconventional expectation value, 
when the density operator commutes with the Hamiltonian and the conserved quantities.
The simple extensions of the density operator for the GGE are shown in quantum statistics.

This  paper is organized as follows. 
In section \ref{sec:quantanal}, the quantum analysis is briefly reviewed, and the variations of some functionals are calculated. 
In section \ref{density:operator:mep}, the equation of the density operator is derived in the MEP.
The density operators for the von Neumann, Tsallis, and R\'enyi entropies are derived when conserved quantities exist.
The last section is assigned for discussion and conclusion.


\section{Quantum analysis and variations of functionals}
\label{sec:quantanal}
\subsection{Basics of quantum analysis}
We begin with the following differential $df(\qmA)$ with operators $\qmA$ and $d\qmA$:
\begin{align}
df(\qmA) = \lim_{\varepsilon \rightarrow 0} \frac{f(\qmA + \varepsilon d\qmA) -f(\qmA)}{\varepsilon} .
\end{align}
The operators $\qmA$ and $d\qmA$ do not commute generally. 
Symbolically, the above differential is represented as 
\begin{align}
df(\qmA) = \frac{df(\qmA)}{d\qmA} d\qmA . 
\end{align}
The hyperoperator $df(\qmA)/d\qmA$ maps  $d\qmA$ to $df(\qmA)$.  
This hyperoperator $df(\qmA)/d\qmA$ is named quantum derivative in the quantum analysis. 
A hyperoperator $\hL{\qmA}$ is defined as the left multiplication to the operator: 
\begin{align}
\hL{\qmA} \qmB := \qmA \qmB.
\end{align}
The inner derivative $\hdelta{\qmA}$ is defined by using the commutation relation:
\begin{align}
\hdelta{\qmA} \qmB := [\qmA, \qmB] = \qmA \qmB - \qmB \qmA. 
\end{align}
The hyperoperator $\hL{\qmA}$ and the inner derivative $\hdelta{\qmA}$ commute:
\begin{align}
\hL{\qmA} \hdelta{\qmA} \qmB = \hdelta{\qmA} \hL{\qmA} \qmB .
\end{align}
This property is useful in calculations.
Hereafter, we do not distinguish between $\hL{\qmA}$ and $\qmA$ throughout this paper as in Ref. \cite{Suzuki-commun-math}

The quantum derivative $df(\qmA)/d\qmA$ is represented as follows:
\begin{align}
\frac{df(\qmA)}{d\qmA} = \int_0^1 \ dt f^{(1)} ( \qmA - t \hdelta{\qmA}), 
\label{rep:quant-derivative}
\end{align}
where $f^{(k)}(x)$ is the $k$-th derivative of $f(x)$ with respect to a classical variable $x$.
The first order quantum Taylor expansion for operators $\qmA$ and $\qmB$ is represented with $df(\qmA)/d\qmA$:
\begin{align}
f(\qmA + x \qmB) = f(\qmA) + x \frac{df(\qmA)}{d\qmA} \qmB + O(x^2). 
\label{rep:quant-taylor}
\end{align}

The variation of the functional is often used. 
The variation of the functional $F[A(t)]$ of $A(t)$ with a variable $t$ \cite{Suzuki:JMathPhys, Suzuki:Book:QuantAnalysis} is defined as follows:
\begin{align}
\delta F[\qmA(t)] := \lim_{\varepsilon \rightarrow 0} \frac{F[\qmA(t)+\epsilon (\delta \qmA(t))] - F[\qmA(t)]}{\varepsilon} . 
\end{align}

These expressions are used to derive the density operator in the MEP.
We attempt to obtain the expressions of variations for some functionals in the next subsection.

\subsection{Variations of functionals with quantum analysis}
In this subsection, we calculate variations of some functionals with the quantum analysis. 
We treat following functionals: 
\begin{subequations}
\begin{align}
&F(\qmrho) = \Tr(f(\qmrho)),\\
&G(\qmrho; \qmA) = \frac{\Tr(g(\qmrho) \qmA)}{\Tr(g(\qmrho))},\\
&H(\qmrho) = h(F(\qmrho)) \equiv h(\Tr(f(\qmrho))). 
\end{align}
\end{subequations}
First, we treat the functional $F(\qmrho)$. 
The variation  of $F(\qmrho)$ is obtained by applying the quantum Taylor expansion, Eq.~\eqref{rep:quant-taylor}:
\begin{align}
\delta F(\qmrho) 
&= \lim_{\varepsilon \rightarrow 0} \varepsilon^{-1} 
\bigg\{ \Tr [ f(\qmrho + \varepsilon(\delta \qmrho))] -  \Tr[ f(\qmrho) ] \bigg\}
\nonumber \\
&=\lim_{\varepsilon \rightarrow 0} \varepsilon^{-1}  \bigg\{
\Tr \left[ f(\qmrho) + \frac{df(\qmrho)}{d\qmrho} \varepsilon (\delta \qmrho) + O(\varepsilon^2)  \right]  - \Tr[ f(\qmrho) ] 
\bigg\}
\nonumber \\
&= \Tr \left[ \frac{df(\qmrho)}{d\qmrho}  (\delta \qmrho) \right] . 
\end{align}
Applying Eq.~\eqref{rep:quant-derivative} to the above equation,  we obtain
\begin{align}
\delta F(\qmrho) 
&= \Tr \left[ \int_0^1 dt f^{(1)}(\qmrho - t \hdelta{\qmrho})  (\delta \qmrho) \right]
= \Tr \left[ \int_0^1 dt  \sum_{k=0}^{\infty} \frac{1}{k!} f^{(k+1)}(\qmrho) (- t)^k \big( (\hdelta{\qmrho})^k   (\delta \qmrho) \big) \right]
\nonumber \\
&= \sum_{k=0}^{\infty} \frac{(-1)^k}{(k+1)!} \ \Tr \left[ f^{(k+1)}(\qmrho)  \big( (\hdelta{\qmrho})^k   (\delta \qmrho) \big) \right] .
\label{deltaF:tmp}
\end{align}
We focus on the trace in Eq.~\eqref{deltaF:tmp} for $k \ge 1$.
The term $((\hdelta{\qmrho})^k (\delta \qmrho))$ generates $2^k$ terms of the form $\qmrho^j (\delta \qmrho) \qmrho^{k-j}$ $(0 \le j  \le k)$ . 
This term gives the same contribution when the cyclic permutation property of trace is hold:
\begin{align}
\Tr \left[ f^{(k+1)}(\qmrho)  \qmrho^j (\delta \qmrho) \qmrho^{k-j} \right] 
= \Tr \left[  \qmrho^{k-j} f^{(k+1)}(\qmrho)  \qmrho^j (\delta \qmrho) \right] 
= \Tr \left[  \ f^{(k+1)}(\qmrho)  \qmrho^k (\delta \qmrho) \right] .
\end{align}
The number of the terms with plus sign equals that with minus sign. 
Therefore, we obtain 
\begin{align}
\Tr \left[ f^{(k+1)}(\qmrho)  \big( (\hdelta{\qmrho})^k   (\delta \qmrho) \big) \right] =  0 \qquad (k \ge 1). 
\label{eqn:trace:expanded}
\end{align}
Equation~\eqref{eqn:trace:expanded} is also proved with the result given in \ref{sec:appendix:A}.
Equation~\eqref{eqn:trace:expanded} gives 
\begin{align}
\delta F(\qmrho) = \Tr \left[ f^{(1)}(\qmrho)  (\delta \qmrho) \right] .
\end{align}

Next, we attempt to calculate $\delta G(\qmrho; \qmA)$. 
We pay attention that $\qmrho$ and $\qmA$ do not commute generally. 
In the same way, we obtain the expression of $\delta G(\qmrho; \qmA)$ by applying Eqs.~\eqref{rep:quant-derivative} and \eqref{rep:quant-taylor}.
\begin{align}
\delta G(\qmrho; \qmA)
&= 
\frac{1}{\Tr (g(\qmrho))} \Bigg[ 
\displaystyle\sum_{k=0}^{\infty} \frac{(-)^k}{(k+1)!}  \Tr \Big( g^{(k+1)} (\qmrho)\Big( (\hdelta{\qmrho})^k (\delta\qmrho) \Big) \qmA \Big) 
\Bigg]
\nonumber \\ & \qquad 
- 
\frac{\Tr(g(\qmrho) \qmA)}{\Tr(g(\qmrho))}
\frac{1}{ \Tr (g(\qmrho) )} \Bigg[ \Tr \Big( g^{(1)} (\qmrho) (\delta\qmrho) \Big)  \Bigg] . 
\label{eqn:deltaG}
\end{align}

Equation~\eqref{eqn:deltaG} is simplified when the conventional expectation value is employed: the function $g(x)$ is  $cx$, where $c$ is a constant. 
Therefore, we have $g^{(1)}(x)  = c$ and  $g^{(j)}(x) = 0$ $(j \ge 2)$. 
This gives 
\begin{align}
\delta G(\qmrho; \qmA)
&= 
\frac{\Tr ((\delta \qmrho) \qmA)}{\Tr (\qmrho)} 
-  \frac{\Tr(\qmrho\qmA)}{\Tr(\qmrho)} \frac{\Tr(\delta\qmrho)}{ \Tr (\qmrho)}  
\qquad \mathrm{for\ } g(\qmrho) = c \qmrho.
\label{deltaG:conventional}
\end{align}
We note that the commutation relation between $\qmrho$ and $\qmA$ is not assumed to obtain Eq.~\eqref{deltaG:conventional}. 
Equation~\eqref{eqn:deltaG} is also simplified 
when the commutation relation between $\qmrho$ and $\qmA$ is a function of $\qmrho$: $[\qmrho,\qmA] = r(\qmrho)$.
The trace, $\Tr( u(\qmrho) ( (\hdelta{\qmrho})^k (\delta\qmrho) ) \qmA )$, is 
\begin{align}
\Tr( u(\qmrho) ( (\hdelta{\qmrho})^k (\delta\qmrho) ) \qmA ) = 
\left\{
\begin{array}{ll}
0&(k \ge 2)\\
- \Tr(u(\qmrho) r(\qmrho) (\delta \qmrho)) & (k =1)\\
\Tr(u(\qmrho) (\delta \qmrho) \qmA) & (k =0)
\end{array}
\right.
, 
\end{align}
where $u(\qmrho)$ is a function of $\qmrho$ (see \ref{sec:appendix:B}).
We have
\begin{align}
\delta G(\qmrho; \qmA)
&= 
\frac{1}{ \Tr (g(\qmrho))}  
\Tr \Big\{ \Big[ ( \qmA -  G(\qmrho; \qmA) ) g^{(1)}(\qmrho)  + \frac{1}{2} g^{(2)}(\qmrho) r(\qmrho) \Big] (\delta \qmrho) \Big\}
\qquad \mathrm{for\ } [\qmrho, \qmA] = r(\qmrho). 
\label{eqn:deltaG:rp}
\end{align}

Finally, we treat $H(\qmrho)$. 
The variation $\delta H(\qmrho)$ is easily obtained by using the expression of $\delta F(\qmrho)$. 
\begin{align}
\delta H(\qmrho) 
&= \lim_{\varepsilon \rightarrow 0} \frac{H(\qmrho+\varepsilon (\delta \qmrho)) - H(\qmrho)}{\varepsilon}
= \lim_{\varepsilon \rightarrow 0} \frac{h(F(\qmrho) +\varepsilon \delta F(\qmrho)  + O(\varepsilon^2)) - h(F(\qmrho))}{\varepsilon}
\nonumber \\
&= \lim_{\varepsilon \rightarrow 0} \frac{[h(F(\qmrho)) + h^{(1)} (F(\qmrho)) ( \varepsilon \delta F(\qmrho)  + O(\varepsilon^2)) + O(\varepsilon^2)]- h(F(\qmrho))}{\varepsilon}
\nonumber \\
&=  h^{(1)} (F(\qmrho)) (\delta F(\qmrho) ) .
\end{align}

In the next section, these results are used to derive the density operators in the MEP.

\section{Derivation of the density operators in the maximum entropy principle}
\label{density:operator:mep}

\subsection{Derivation of the equation of the density operator with the quantum analysis}
We treat the entropy $S(\qmrho)$ under some constraints, where $\qmrho$ is a density operator. 
The expectation value of a quantity $\hat{A}$ is denoted by $\<\hat{A}>$.
The quantity  $\qmQa$ is the operator of a conserved quantity. 
Therefore, the operator $\qmQa$ and the Hamiltonian $\qmH$  commute: $[\qmH, \qmQa] = 0$.
The number of the constraints represented by the operators $\qmQa$ is $M$.
Therefore, we treat the variational problem with the following functional $I(\qmrho)$:
\begin{align}
I(\qmrho) &:= S(\qmrho) - \alpha (\Tr \qmrho  - 1) - \beta (\<\qmH > - E) - \sum_{a=1}^{M} \gamma^{[a]} (\<\qmQa> - \clQa) , 
\end{align}
where $\alpha, \beta, \gamma^{[a]}$ are Lagrange multipliers, $E$ is the energy, 
and the expectation value of the physical quantity $\qmQa$ is $\clQa$.
We note that the above expectation value is not always the conventional expectation value defined by $\Tr(\qmrho \qmA)/\Tr(\qmrho)$.

The  maximum entropy principle requires $\delta I(\qmrho) =0$. 
We treat the following entropy and average in this paper. 
\begin{subequations}
\begin{align}
&S(\qmrho) := H(\qmrho) \equiv h(\Tr(f(\qmrho)))  = h(F(\qmrho)), \\
&\<\qmQa> := G(\qmrho; \qmQa) \equiv \frac{\Tr(g(\qmrho) \qmQa)}{\Tr(g(\qmrho))} . 
\end{align}
\end{subequations}
The variation of the entropy $\delta S(\qmrho)$ is given by 
\begin{align}
\delta S(\qmrho) 
= h^{(1)}(F(\qmrho)) \Tr \big( f^{(1)} (\qmrho) (\delta\qmrho) \big)
= \Tr \big(  h^{(1)}(F(\qmrho)) f^{(1)} (\qmrho) (\delta\qmrho) \big) .
\label{eqn:deltaS}
\end{align}
The functional $I(\qmrho)$ is rewritten with $G(\qmrho, \qmA)$: 
\begin{align}
I(\qmrho)  = S(\qmrho) - \alpha (\Tr \qmrho  - 1) - \beta (G(\qmrho; \qmH) - E) - \sum_{a=1}^{M} \gamma^{[a]} (G(\qmrho; \qmQa) - \clQa) .
\end{align}
The variation of $I(\qmrho)$ is given by 
\begin{align}
\delta I(\qmrho)  
&= \delta S(\qmrho) - \alpha (\Tr (\delta \qmrho)) - \beta (\delta G(\qmrho; \qmH))  - \sum_{a=1}^{M} \gamma^{[a]} (\delta G(\qmrho; \qmQa)) .
\label{eqn:deltaI2}
\end{align}
With this expression, we attempt to find the density operators in the following subsection.

\subsection{Density operators in particular cases}
The variation $\delta I(\qmrho)$, Eq.~\eqref{eqn:deltaI2},  is reduced to simple expressions in particular cases. 
In this subsection, we treat two cases: 
(i) the conventional expectation value is employed, and
(ii) the commutation relations, $[\qmrho, \qmH]=r^{[0]}(\qmrho)$ and $[\qmrho, \qmQa]=r^{[a]}(\qmrho)$ $(a=1,2,\cdots, M)$,
are satisfied. 
Especially, density operators are given for $r^{[0]}(\qmrho) = r^{[a]}(\qmrho) = 0$ when the escort average is employed.

\subsubsection{The cases of conventional expectation value}
The function $g(x)$ is $cx$ for the conventional expectation value, where $c$ is a constant. 
This leads to $g^{(1)} = c$ and  $g^{(k)}(x) = 0$ $(k \ge 2)$.
Equation~\eqref{eqn:deltaI2} is reduced to the following equation:
\begin{align}
\delta I(\qmrho) 
&=  \Tr \Bigg[ \Bigg( h^{(1)}(F(\qmrho))  f^{(1)}(\qmrho)  - \alpha  - \beta \Bigg( \frac{ ( \qmH - \< \qmH> )}{\Tr( \qmrho)} \Bigg) 
- \sum_{a=1}^{M} \gamma^{[a]} \Bigg( \frac{ (\qmQa - \<\qmQa>)}{\Tr (\qmrho)} \Bigg)
\Bigg) (\delta \qmrho)  \Bigg] .
\end{align}
The requirement $\delta I(\qmrho)=0$ leads to 
\begin{align}
 h^{(1)}(F(\qmrho))  f^{(1)}(\qmrho)  - \alpha  - \beta \Bigg( \frac{ ( \qmH - \< \qmH> )}{\Tr( \qmrho)} \Bigg) 
- \sum_{a=1}^{M} \gamma^{[a]} \Bigg(\frac{ (\qmQa - \<\qmQa>)}{\Tr (\qmrho)} \Bigg)
=0 .
\end{align}
Considering the requirement $\Tr (\qmrho) = 1$, we have 
\begin{align}
 h^{(1)}(F(\qmrho))  f^{(1)}(\qmrho)  
= \alpha  + \beta  ( \qmH - \< \qmH> ) + \sum_{a=1}^{M} \gamma^{[a]}  (\qmQa - \<\qmQa> ) .
\label{eqn:qmrho:conventinal:general}
\end{align}
It is significant that Eq.~\eqref{eqn:qmrho:conventinal:general} is derived 
without assuming the commutation relations between operators: the relation between $\qmQa$ and $\qmQb$ is not assumed.
Equation~\eqref{eqn:qmrho:conventinal:general} indicates that 
the argument of the density operator for the canonical ensemble is simply extended to the argument for the GGE. 

First, we treat the von Neumann entropy as an example. 
The function $f(x)$ is $-x\ln x$ and $h(x)$ is $x$ for the von Neumann entropy.
The derivatives, $f^{(1)}(x) =-\ln x -1$ and $h^{(1)}(x) = 1$, are obtained. 
Inserting these expressions into Eq.~\eqref{eqn:qmrho:conventinal:general}, we have
\begin{align}
\qmrho = \exp \Big( - \alpha  -1 - \beta  ( \qmH - \< \qmH> ) - \sum_{a=1}^{M} \gamma^{[a]}  (\qmQa - \<\qmQa> )\Big) .
\end{align}
The requirement $\Tr (\qmrho) = 1$ leads to 
\begin{subequations}
\begin{align}
& \qmrho = \frac{1}{Z_{\mathrm{N}}^{\mathrm{conv}}} \exp \Bigg(  - \beta \qmH - \displaystyle\sum_{a=1}^{M} \gamma^{[a]} \qmQa \Bigg) ,\\
& Z_{\mathrm{N}}^{\mathrm{conv}} = \Tr \Bigg[ \exp \Bigg(- \beta \qmH - \displaystyle\sum_{a=1}^{M} \gamma^{[a]} \qmQa \Bigg) \Bigg] . 
\end{align}
\end{subequations}
This is just the density operator for the GGE.

Next, we treat the Tsallis entropy.
The function $f(x)$ is $x^q$ and $h(x)$ is $(1-x)/(q-1)$ for the Tsallis entropy.
The derivatives, $f^{(1)}(x) = q x^{q-1}$ and $h^{(1)}(x) = 1/(1-q)$, are inserted into Eq.~\eqref{eqn:qmrho:conventinal:general}:
\begin{align}
\frac{q}{1-q} \qmrho^{q-1} = \alpha  + \beta  ( \qmH - \< \qmH> ) + \sum_{a=1}^{M} \gamma^{[a]}  (\qmQa - \<\qmQa> ) . 
\label{eqn:diff:rho:conventional}
\end{align}
With the requirement $\Tr (\qmrho) = 1$, we have
\begin{subequations}
\begin{align}
& \qmrho =  \frac{1}{Z_{\mathrm{T}}^{\mathrm{conv}} }
\Big( 1 +  \tilde{\beta}  ( \qmH - \< \qmH> ) + \sum_{a=1}^{M}  \tilde{\gamma}^{[a]}  (\qmQa - \<\qmQa> \Big)^{\frac{1}{(q-1)}} ,
\\ 
& Z_{\mathrm{T}}^{\mathrm{conv}} = 
\Tr \Big[\Big( 1 +  \tilde{\beta}  ( \qmH - \< \qmH> ) + \sum_{a=1}^{M}  \tilde{\gamma}^{[a]}  (\qmQa - \<\qmQa> \Big)^{\frac{1}{(q-1)}} \Big] ,\\
&\tilde{\beta} := \beta/\alpha, \\
&\tilde{\gamma}^{[a]} := \gamma^{[a]}/\alpha .
\end{align}
\label{Tsallis:conventional:2}
\end{subequations}

We have the following relation by multiplying $\qmrho$  and calculating the trace  from Eq.~\eqref{eqn:diff:rho:conventional}:
\begin{align}
\frac{q}{1-q} \Tr (\qmrho^{q}) = \alpha . 
\label{rel:conventional:alpha:rho}
\end{align}

Finally, we treat the R\'enyi entropy.
The function $f(x)$ is $x^q$ and $h(x)$ is $\ln x/(1-q)$ for the R\'enyi entropy.
The derivatives, $f^{(1)} = q x ^{q-1}$ and $h^{(1)}(x) = 1/( (1-q) x)$, are inserted into  Eq.~\eqref{eqn:qmrho:conventinal:general}:
\begin{align}
\frac{q}{1-q} \frac{1}{\Tr(\qmrho^q)} \qmrho^{q-1} = \alpha  + \beta  ( \qmH - \< \qmH> ) + \sum_{a=1}^{M} \gamma^{[a]}  (\qmQa - \<\qmQa> ) . 
\label{eqn:Renyidiff:rho:conventional}
\end{align}
With the requirement $\Tr (\qmrho) = 1$, we have
\begin{subequations}
\begin{align}
& \qmrho =  \frac{1}{Z_{\mathrm{R}}^{\mathrm{conv}} }
\Big( 1 +  \tilde{\beta}  ( \qmH - \< \qmH> ) + \sum_{a=1}^{M}  \tilde{\gamma}^{[a]}  (\qmQa - \<\qmQa> \Big)^{\frac{1}{(q-1)}} , 
\\ 
& Z_{\mathrm{R}}^{\mathrm{conv}} = 
\Tr \Big[\Big( 1 +  \tilde{\beta}  ( \qmH - \< \qmH> ) + \sum_{a=1}^{M}  \tilde{\gamma}^{[a]}  (\qmQa - \<\qmQa> \Big)^{\frac{1}{(q-1)}} \Big]  ,\\
&\tilde{\beta} := \beta/\alpha, \\
&\tilde{\gamma}^{[a]} := \gamma^{[a]}/\alpha .
\end{align}
\label{Renyi:conventional:2}
\end{subequations}
We have the following relation by multiplying $\qmrho$ and calculating the trace from Eq.~\eqref{eqn:Renyidiff:rho:conventional}:
\begin{align}
\frac{q}{1-q} = \alpha . 
\end{align}

The density operator for the canonical ensemble is simply extended to
the density operator for the GGE in the conventional expectation value case.

\subsubsection{The cases of $[\qmrho, \qmH]=r^{[0]}(\qmrho)$ and $[\qmrho, \qmQa]=r^{[a]}(\qmrho)$ $(a=1,2,\cdots, M)$}

We treat the cases of $[\qmrho, \qmH]=r^{[0]}(\qmrho)$ and $[\qmrho, \qmQa]=r^{[a]}(\qmrho)$ $(a=1,2,\cdots, M)$, 
where $r^{[0]}(\qmrho)$ and $r^{[a]}(\qmrho)$ are functions of $\qmrho$. 
The cases contain the following conditions: $[\qmrho, \qmH]=0$ and $[\qmrho, \qmQa]=0$ $(a=1,2,\cdots, M)$. 
The density operator constructed from $\qmH$ and $\qmQa$ satisfies $[\qmrho, \qmH]=[\qmrho, \qmQa]=0$
when the commutation relations $[\qmQa, \qmQb]=0$ $(a, b=1,2,\cdots, M)$ are satisfied.

The variation $\delta I(\qmrho)$ is given by using Eqs.~\eqref{eqn:deltaS} and \eqref{eqn:deltaG:rp}
in the case of $[\qmrho, \qmH]=r^{[0]}(\qmrho)$ and $[\qmrho, \qmQa]=r^{[a]}(\qmrho)$ $(a=1,2,\cdots, M)$: 
\begin{align}
\delta I(\qmrho)  
= & 
\Tr \Bigg\{ \Bigg[  h^{(1)}(F(\qmrho))  f^{(1)}(\qmrho)  - \alpha  
- \frac{\beta}{\Tr( g(\qmrho))}  \Big( \big( \qmH - \< \qmH> \big) g^{(1)}(\qmrho)  + \frac{1}{2} g^{(2)}(\qmrho) r^{[0]}(\qmrho) \Big)
\nonumber \\ & \quad
- \sum_{a=1}^{M}  \frac{\gamma^{[a]}}{\Tr (g(\qmrho))} \Big( \big(\qmQa - \<\qmQa> \big) g^{(1)} (\qmrho) + \frac{1}{2} g^{(2)}(\qmrho) r^{[a]}(\qmrho) \Big) 
\Bigg] 
(\delta \qmrho)  \Bigg\} .
\end{align}
Therefore, the condition $\delta I(\qmrho) = 0$ gives 
\begin{align}
h^{(1)} (F(\qmrho))  f^{(1)}(\qmrho)  - \alpha  
& - \frac{\beta}{\Tr( g(\qmrho))}  \Big( \big( \qmH - \< \qmH> \big) g^{(1)}(\qmrho)  + \frac{1}{2} g^{(2)}(\qmrho) r^{[0]}(\qmrho) \Big)
\nonumber \\ & 
- \sum_{a=1}^{M}  \frac{\gamma^{[a]}}{\Tr (g(\qmrho))} \Big( \big(\qmQa - \<\qmQa> \big) g^{(1)} (\qmrho) + \frac{1}{2} g^{(2)}(\qmrho) r^{[a]}(\qmrho) \Big) 
= 0 .
\end{align}

We treat the cases of $[\qmrho, \qmH] = [\qmrho, \qmQa] =0$ $(a=1, 2, \cdots, M)$ hereafter.
The requirement $\delta I(\qmrho) =0$ for $[\qmrho, \qmH] = [\qmrho,\qmQa] = 0$ gives
\begin{align}
& 
h^{(1)}(F(\qmrho))  f^{(1)}(\qmrho)  - \alpha  - \frac{\beta}{\Tr( g(\qmrho))} \big( \qmH - \< \qmH> \big) g^{(1)}(\qmrho) 
\nonumber \\ & \qquad 
- \sum_{a=1}^{M} \frac{\gamma^{[a]}}{\Tr (g(\qmrho))} \big(\qmQa-\<\qmQa> \big) g^{(1)} (\qmrho) 
= 0 .
\label{eqn:rho-Q=0}
\end{align}
With Eq.~\eqref{eqn:rho-Q=0}, 
we attempt to find the density operators for the Tsallis and R\'enyi entropies with the escort average.

We attempt to obtain the density operator for the Tsallis entropy with the escort average.
The function $f(x)$ is $x^q$, $h(x)$ is $(1-x)/(q-1)$ for the Tsallis entropy, and the function $g(x)$ is $x^q$ for the escort average.
Inserting these functions into Eq.~\eqref{eqn:rho-Q=0}, we have
\begin{align}
\Bigg( \frac{q}{1-q}  \Bigg) \qmrho^{q-1}   - \alpha  
- q\beta \Bigg( \frac{( \qmH - \< \qmH> )}{\Tr( \qmrho^q)} \Bigg) \qmrho^{q-1} 
- \sum_{a=1}^{M} q \gamma^{[a]} \Bigg( \frac{(\qmQa - \<\qmQa>)}{ \Tr (\qmrho^q)} \Bigg) \qmrho^{q-1} 
=0 . 
\label{eqn:Tsallis:escort}
\end{align}
Multiplying $\qmrho$ and taking the trace, we obtain
\begin{align}
\frac{q}{1-q} \Tr \qmrho^q - \alpha = 0. 
\label{alpha:Tr}
\end{align}

We have the density operator for the Tsallis entropy with the escort average  with the requirement $\Tr \qmrho =1$. 
\begin{subequations}
\begin{alignat}{2}
&\ \qmrho &&= \frac{1}{Z^{\mathrm{esc}}_{\mathrm{T}}}
\Bigg(
1- (1-q) \beta \Bigg( \frac{( \qmH - \< \qmH> )}{\Tr( \qmrho^q)} \Bigg) 
- \sum_{a=1}^{M} (1-q)  \gamma^{[a]} \Bigg( \frac{(\qmQa - \<\qmQa>)}{ \Tr (\qmrho^q)} \Bigg)
\Bigg)^{\frac{1}{1-q}} , 
\label{Tsallis:escort:rho} \\ 
& Z^{\mathrm{esc}}_{\mathrm{T}} & & = 
\Tr \Bigg[ \Bigg(
1- (1-q) \beta \Bigg( \frac{( \qmH - \< \qmH> )}{\Tr( \qmrho^q)} \Bigg) 
- \sum_{a=1}^{M} (1-q)  \gamma^{[a]} \Bigg( \frac{(\qmQa - \<\qmQa>)}{ \Tr (\qmrho^q)} \Bigg)
\Bigg)^{\frac{1}{1-q}} \Bigg] . 
\end{alignat}
\end{subequations}
Equation~\eqref{Tsallis:escort:rho} is the density operator for the GGE for the Tsallis entropy with the escort average. 
The conditions $[\qmrho, \qmQa] =0$ $(a=1, 2, \cdots, M)$ are satisfied in the case of $[\qmQa, \qmQb]=0$.
Therefore, 
Eq.~\eqref{Tsallis:escort:rho} gives the density operator for the Tsallis entropy with the escort average 
when the conserved quantities commute: $[\qmQa, \qmQb]=0$.

Another example is the R\'enyi entropy with the escort average. 
For the R\'enyi entropy, the function $f (x)$ is $x^q$, $g(x)$ is $x^q$, and $h(x)$ is $(\ln x)/(1 - q)$.
These functions yield 
\begin{align}
\left( \frac{q}{1-q} \frac{1}{\Tr(\qmrho^q)}  \right)  \qmrho^{q-1} 
-\alpha - q \beta \Bigg( \frac{(\qmH - \<\qmH>) }{\Tr(\qmrho^q)} \Bigg) \qmrho^{q-1} 
- \sum_{a=1}^{M} q \gamma^{[a]} \Bigg( \frac{\qmQa - \<\qmQa>}{\Tr(\qmrho^{q})} \Bigg) \qmrho^{q-1}= 0  . 
\label{eqn:Renyi:tmp-eq}
\end{align}
Multiplying $\qmrho$ and taking the trace, we obtain
\begin{align}
\left( \frac{q}{1-q}  \right) -\alpha  = 0 .
\label{eqn:Renyi:alpha-q}
\end{align}

We have the density operator for the R\'enyi entropy with the escort average with the requirement $\Tr \qmrho = 1$.
\begin{subequations}
\begin{align}
& \qmrho =  \frac{1}{Z^{\mathrm{esc}}_{\mathrm{R}}} 
\Bigg( 1 - (1-q) \beta (\qmH - \<\qmH>) - \sum_{a=1}^{M} (1-q)  \gamma^{[a]} (\qmQa - \<\qmQa>) \Bigg)^{\frac{1}{1-q}} ,
\label{Renyi:escort:rho} \\
& Z^{\mathrm{esc}}_{\mathrm{R}} = \Tr \Bigg[ \Bigg( 1 - (1-q) \beta (\qmH - \<\qmH>) - \sum_{a=1}^{M} (1-q)  \gamma^{[a]} (\qmQa - \<\qmQa>) \Bigg)^{\frac{1}{1-q}}   \Bigg] . 
\end{align}
\end{subequations}
Equation~\eqref{Renyi:escort:rho} is the density operator for the GGE for the R\'enyi entropy with the escort average. 
The conditions $[\qmrho, \qmQa] =0$ $(a=1, 2, \cdots, M)$ are satisfied for $[\qmQa, \qmQb]=0$.
Eq.~\eqref{Renyi:escort:rho} gives the density operator for the R\'enyi entropy with the escort average 
when the conserved quantities commute: $[\qmQa, \qmQb]=0$. 
The density operator for the R\'enyi entropy is similar to the density operator for the Tsallis entropy.

\section{Discussion and Conclusion}
We derived the equation of the density operator in the maximum entropy principle 
by using the quantum analysis.
The derived equation is simplified in particular cases, 
and the density operators for the generalized Gibbs ensemble are derived: 
the density operators for the von Neumann entropy, the Tsallis entropy, and the R\'enyi entropy
in the case of the conventional expectation value  
and the density operators for the Tsallis entropy and the R\'enyi entropy
in the case of the escort average (the normalized $q$-expectation value). 
The obtained density operators are the simple extensions of the well-known density operators for the canonical ensemble.

The derived equation of the density operator is simplified in two cases in the present paper. 
The equation is simplified when the conventional expectation value is employed
and the equation is also simplified 
when the following commutation relations are satisfied: 
$[\qmrho, \qmH] = r^{[0]}(\qmrho)$ and $[\qmrho, \qmQa] = r^{[a]}(\qmrho)$ $(a=1,2,\cdots, M)$, 
where $\qmrho$ is the density operator, $\qmH$ is the Hamiltonian, $\qmQa$ is the conserved quantity, 
and $r^{[j]}(\qmrho)$ is a function of $\qmrho$ $(j=0, 1, 2, \cdots, M)$.

The argument of the density operator for the generalized Gibbs ensemble is given
by adding the conserved quantities to the argument of the density operator for the canonical ensemble
when the conventional expectation value is employed:
the argument $-\beta (\qmH - \<\qmH>)$ in the canonical ensemble is replaced with 
the argument $-\beta (\qmH - \<\qmH>) - \displaystyle\sum_a \gamma^{[a]} (\qmQa - \<\qmQa>) $ in the generalized Gibbs ensemble.
There is no assumption of the commutation relations between the conserved quantities, 
while the Hamiltonian and the conserved quantity commute.

The argument of the density operator for the canonical ensemble is extended simply
to the argument of the density operator for the generalized Gibbs ensemble
in the case of the escort average (the normalized $q$-expectation value),
as found in the case of the conventional expectation value, 
when the density operator $\qmrho$, the Hamiltonian $\qmH$,
and conserved quantities $\qmQa$ ($a = 1, 2, \cdots, M$) satisfy the following commutation relations:
$[\qmrho, \qmH] = 0$ and $[\qmrho, \qmQa]=0$.
We note that the conditions, $[\qmrho, \qmH] = 0$ and $[\qmrho, \qmQa]=0$, are satisfied
when the commutation relations $[\qmQa, \qmQb]=0$ ($a, b = 1, 2, \cdots, M$) are satisfied. 
The argument of the density operator is extended simply even when the expectation value is not conventional.

It is clearly shown for various entropies that 
the argument of the density operator for the canonical ensemble is simply extended to the argument for the generalized Gibbs ensemble 
when the conventional expectation value is employed, even though the conserved quantities do not commute.
This implies that the well-known density operator for the generalized Gibbs ensemble appears when there are conserved quantities.   
In contrast,
the argument of the density operator for the canonical ensemble is not always extended simply
in the case of the unconventional expectation value,
when there are conserved quantities. 
The argument of the density operator may be simply extended in the case of unconventional expectation value 
when additional conditions are imposed. 
Such conditions are given in the present paper:
the density operator and the Hamiltonian commute, and the density operator and the conserved quantities commute. 
The conditions will be satisfied when the conserved quantities commute each other.

We  derived the density operators for the generalized Gibbs ensemble in the particular cases. 
It is not enough to study the derivation of the density operator
in the cases of unconventional expectation values, when the density operator and the conserved quantity do not commute.
The explicit commutation relations between the conserved quantities should be required to derive the density operator. 
The derivation of the density operator in such cases will be studied in the future.

\appendix
\def\hA{\hat{A}} 
\def\hB{\hat{B}} 

\section{Expansion of $((\hdelta{\qmrho})^k \hA)$ and proof of $\Tr ( f(\qmrho) ( (\hdelta{\qmrho})^k \hA ) \hB) = 0$ for $k\ge 1$ in the case of $[\qmrho, \hA] = 0$ and/or $[\qmrho, \hB] = 0$}
\label{sec:appendix:A}
In this appendix, we attempt to expand $((\hdelta{\qmrho})^k \hA)$
and to show the following identity explicitly when $[\qmrho, \qmA] = 0$ and/or $[\qmrho, \qmB] = 0$ are hold: 
 \begin{align}
\Tr \Big( f(\qmrho)\Big( (\hdelta{\qmrho})^k \hA \Big) \hB \Big) = 0 \qquad (k \ge 1) .  
\end{align}

First, we expand the quantity $((\hdelta{\qmrho})^k \hA)$ by considering the definition of $\hdelta{\rho}$: $\hdelta{\qmrho} \hA = \qmrho \hA - \hA \qmrho$.
\begin{align}
( (\hdelta{\qmrho})^k \hA ) = \sum_{j=0}^{k} C_j^{(k)} \qmrho^j \hA \qmrho^{k-j}.
\label{eqn:expansion:hdeltahA}
\end{align}
We attempt to find the recursion formula of $C_j^{(k)}$ to obtain the expression of $C_j^{(k)}$.
By applying $\hdelta{\qmrho}$ to $(\hdelta{\qmrho})^{k-1} \hA$, we have 
\begin{subequations}
\begin{align}
&C_{j}^{(k)} = C_{j-1}^{(k-1)} -  C_j^{(k-1)} \qquad  (1 \le j \le k-1),\\
&C_{k}^{(k)} = C_{k-1}^{(k-1)}, \\
&C_{0}^{(k)} = - C_{0}^{(k-1)}.
\end{align}
\end{subequations}

Next, we expand $(x-y)^k$ as 
\begin{align}
(x-y)^k = \sum_{j=0}^{k} D_j^{(k)} x^j y^{k-j}, \qquad  D_j^{(k)}  =  (-1)^{k-j} \left( \begin{array}{c} k \\ j \end{array} \right).
\label{expansion:x-y}
\end{align}
By applying $(x-y)$ to $(x-y)^{k-1}$, we obtain
\begin{subequations}
\begin{align}
&D_{j}^{(k)} = D_{j-1}^{(k-1)} -  D_j^{(k-1)} \qquad  (1 \le j \le k-1), \\
&D_{k}^{(k)} = D_{k-1}^{(k-1)}, \\
&D_{0}^{(k)} = - D_{0}^{(k-1)}. 
\end{align}
\end{subequations}
The recursion formula of $C_j^{(k)}$ is equivalent to that of  $D_j^{(k)}$. 
The initial conditions are given by
\begin{subequations}
\begin{align} 
C_0^{(1)} = -1, \ C_1^{(1)} = 1,\\
D_0^{(1)} = -1, \ D_1^{(1)} = 1.
\end{align} 
\end{subequations}
Therefore, $C_j^{(k)}$ equals $D_j^{(k)}$. 
We obtain
\begin{align}
C_j^{(k)} = D_j^{(k)} =  (-1)^{k-j} \left( \begin{array}{c} k \\ j \end{array} \right) . 
\label{equivalence}
\end{align}
We have the expansion of $( (\hdelta{\qmrho})^k \hA)$, Eq.~\eqref{eqn:expansion:hdeltahA} with Eq.~\eqref{equivalence}.

With the above result, we attempt to prove the equation: 
\begin{align}
\Tr \Big( f(\qmrho)\Big( (\hdelta{\qmrho})^k \hA \Big) \hB \Big) = 0 \qquad (k \ge 1) .
\end{align}
When the cyclic permutation property of trace is hold and 
the commutation relation $[\qmrho, \hA] = 0$ and/or $[\qmrho, \hB] = 0$ are hold, 
we have the following equation by using Eq.~\eqref{eqn:expansion:hdeltahA}. 
 \begin{align}
&\Tr \Big( f(\qmrho)\Big( (\hdelta{\qmrho})^k \hA \Big) \hB \Big) 
= \sum_{j=0}^{k} C_j^{(k)} \Tr \Big( f(\qmrho)  \qmrho^j \hA \qmrho^{k-j} \hB \Big) 
\nonumber \\ 
&= \sum_{j=0}^{k} C_j^{(k)} \Tr \Big( f(\qmrho)  \qmrho^k \hA \hB \Big) 
= \Tr \Big( f(\qmrho)  \qmrho^k \hA \hB \Big) \sum_{j=0}^{k} C_j^{(k)}  
\qquad (k \ge 1) .  
\end{align}
From Eqs.~\eqref{expansion:x-y} and \eqref{equivalence}, we have  ${\displaystyle\sum_{j=0}^{k} C_j^{(k)}  = 0}$ $(k \ge 1)$. 
We obtain the final result:
 \begin{align}
\Tr \Big( f(\qmrho)\Big( (\hdelta{\qmrho})^k \hA \Big) \hB \Big) = 0
\qquad \mathrm{for}\ [\qmrho, \hA] = 0 \mathrm{\ and/or\ } [\qmrho, \hB] = 0 \qquad  (k \ge 1).  
\label{eqn:trace-eq-zero}
 \end{align}

From the result, Eq.~\eqref{eqn:trace-eq-zero}, 
we have the following identity 
by substituting $\delta \qmrho$ into $\hA$, substituting $\qmQa$ into $\hB$, 
and replacing $f(\qmrho)$ with $g^{(k+1)} (\qmrho)$,  
when $\qmQa$ has the property, $[\qmrho, \qmQa]=0$:
 \begin{align}
\Tr \Big( g^{(k+1)} (\qmrho)\Big( (\hdelta{\qmrho})^k (\delta \qmrho) \Big) \qmQa \Big) = 0 \qquad (k \ge 1 ) .
 \end{align}

The expansion of $((\hdelta{\qmrho})^k \hA)$, Eq.~\eqref{eqn:expansion:hdeltahA}, may be useful to calculate quantities.

\section{Expression of $\Tr{\big( f(\qmrho) ((\hdelta{\qmrho})^k (\delta \qmrho)) \qmB\big)}$ in the case of $[\qmrho, \qmB]=r(\qmrho)$}
\label{sec:appendix:B}

We treat the case of $[\qmrho, \qmB]= \qmrho \qmB - \qmB \qmrho = r(\qmrho)$.
For a function $u(\qmrho)$ and an operator $\qmB$, we have the following equation with the quantum analysis:
\begin{align}
[ u(\qmrho), \qmB] 
= \hdelta{u(\qmrho)}\qmB 
= \frac{\hdelta{u(\qmrho)}}{\hdelta{\qmrho}} \hdelta{\qmrho} \qmB 
= \int_0^1 dt u^{(1)} (\qmrho - t \hdelta{\qmrho}) [\qmrho,\qmB]
= \int_0^1 dt u^{(1)}(\qmrho - t \hdelta{\qmrho}) r(\qmrho) 
.
\end{align}
We have 
\begin{align}
[ u(\qmrho), \qmB] = u^{(1)}(\qmrho) r(\qmrho)  \qquad  \mathrm{for\ } [\qmrho, \qmB] = r(\qmrho) .
\end{align}
Therefore, we obtain the following equation with $[ \qmrho^{k-j}, \qmB] = (k-j) \qmrho^{k-j-1} r(\qmrho)$ 
by substituting $\qmrho^{k-j}$ into $u(\qmrho)$:
\begin{align}
\Tr \Big( f(\qmrho) \qmrho^j (\delta \qmrho) \qmrho^{k-j} \qmB \Big) = 
\Tr \Big( f(\qmrho) \qmrho^k (\delta \qmrho) \qmB \Big) +  (k-j) \Tr \Big( f(\qmrho) r(\qmrho) \qmrho^{k-1} (\delta \qmrho) \Big)
.
\label{calc:trace:temporary}
\end{align}
We have the following equation by using Eq.~\eqref{eqn:expansion:hdeltahA}: 
\begin{align}
\Tr \Big[ f(\qmrho) \big( (\hdelta{\qmrho})^k (\delta \qmrho) \big) \qmB \Big] 
&= \Tr \Big[ f(\qmrho) \big( \sum_{j=0}^k C_j^{(k)} \qmrho^j (\delta \qmrho) \qmrho^{k-j} \big) \qmB \Big] 
\nonumber \\ 
&= \sum_{j=0}^k C_j^{(k)} \Tr \Big[ f(\qmrho) \qmrho^j (\delta \qmrho) \qmrho^{k-j}\qmB \Big] ,
\label{eqn:expansion:B}
\end{align}
where $C_j^{(k)} = (-1)^{k-j} \frac{k!}{j! (k-j)!}$.
Inserting Eq.~\eqref{calc:trace:temporary} into Eq.~\eqref{eqn:expansion:B}, we have
\begin{align}
& \Tr \Big[ f(\qmrho) \big( (\hdelta{\qmrho})^k (\delta \qmrho) \big) \qmB \Big]
\nonumber \\ & \qquad   
= \Tr \Big( f(\qmrho) \qmrho^k (\delta \qmrho) \qmB \Big) \Bigg[ \sum_{j=0}^k C_j^{(k)} \Bigg]
+ \Tr \Big( f(\qmrho) r(\qmrho) \qmrho^{k-1} (\delta \qmrho) \Big) \Bigg[ \sum_{j=0}^{k-1} C_j^{(k)} (k-j) \Bigg]
. 
\end{align}

The sums of $C_j^{(k)}$ satisfy
\begin{subequations}
\begin{align}
&\sum_{j=0}^k C_j^{(k)} = 0       \qquad (k \ge 1) , \\
&\sum_{j=0}^{k-1} C_j^{(k)} (k-j) = 0 \qquad (k \ge 2) . 
\end{align}
\end{subequations}

As a result, we have 
\begin{align}
\Tr \Big[ f(\qmrho) \big( (\hdelta{\qmrho})^k (\delta \qmrho) \big) \qmB \Big] 
= \left\{
\begin{array}{ll}
0 & ( k \ge 2) \\
- \Tr \Big[ f(\qmrho) r(\qmrho) (\delta \qmrho) \Big]  & ( k = 1) \\
\Tr \Big[ f(\qmrho) (\delta \qmrho) \qmB\Big]  & ( k = 0) 
\end{array}
\right.
. 
\end{align}

Again, we have the following result by substituting $\qmA$ into $(\delta \qmrho)$ for $[\qmrho,\qmB]=0$:
\begin{align}
\Tr \Big( f(\qmrho) \big( (\hdelta{\qmrho})^k \qmA \big) \qmB \Big)  = 0,
\qquad \mathrm{for\ } [\qmrho,\qmB]=0 \qquad  ( k \ge 1) .
\end{align}

\end{document}